\documentclass[american,
reprint,
superscriptaddress,
showpacs,
nofootinbib,
amsmath,amssymb,
aps,
pra,
floatfix,
]{revtex4-1}
\usepackage[english]{babel}% Recommended
\usepackage{csquotes}% Recommended
\usepackage{graphicx}% Include figure files
\usepackage{dcolumn}% Align table columns on decimal point
\usepackage[colorlinks=true, citecolor=blue, urlcolor=blue ]{hyperref}% add hypertext capabilities
\usepackage{amsmath,amssymb,amsfonts,amsthm}
\usepackage{mathrsfs}
\usepackage{color}
\usepackage{bbm} %Enables correct fonts for the fields like C,R etc.
\usepackage{bm} %Bold Math
\usepackage{pxfonts,txfonts}
\usepackage{tgpagella}
\usepackage[T1]{fontenc}
\newcommand{\CC}{\mathbbm{C}}

\newcommand\numberthis{\addtocounter{equation}{1}\tag{\theequation}}

\DeclareMathOperator{\tr}{Tr}
\DeclareMathOperator{\diag}{diag}

\DeclareMathOperator{\rank}{rank}

\newcommand{\etal}{\textit{et al.}}

\newcommand{\pal}{\partial}

\newcommand{\ran}{\rangle}
\newcommand{\lan}{\langle}

\newcommand{\id}{\mathbbm{1}}

\newcommand{\norm}[1]{\lVert #1 \rVert}

\newcommand{\N}{\mathcal{N}}

\newtheorem{theorem}{Theorem}
\newtheorem{proposition}[theorem]{Proposition}

\begin{document}

%\preprint{APS/123-QED}

\title{Logarithmic coherence: Operational interpretation of $\ell_1$-norm coherence}

\author{Swapan Rana}
\email{swapanqic@gmail.com} 
\affiliation{ICFO -- Institut de Ciencies Fotoniques, The Barcelona Institute of Science and Technology, 08860 Castelldefels (Barcelona), Spain}

\author{Preeti Parashar}
\affiliation{Physics and Applied Mathematics Unit, Indian Statistical Institute, 203 BT Road, Kolkata, India}

\author{Andreas Winter}
\affiliation{Departament de F\'{\i}sica: Grup d'Informaci\'{o} Qu\`{a}ntica,
	Universitat Aut\`{o}noma de Barcelona, ES-08193 Bellaterra (Barcelona), Spain}
\affiliation{ICREA, Passeig de Llu\'{\i}s Companys, 23, 08010 Barcelona, Spain}

\author{Maciej Lewenstein}
\affiliation{ICFO -- Institut de Ciencies Fotoniques, The Barcelona Institute of Science and Technology, 08860 Castelldefels (Barcelona), Spain}
\affiliation{ICREA, Passeig de Llu\'{\i}s Companys, 23, 08010 Barcelona, Spain}

\date{\today}

\begin{abstract}
We show that the distillable coherence---which is equal to the relative entropy of coherence---
is, up to a constant factor, always bounded by the $\ell_1$-norm measure of coherence 
(defined as the sum of absolute values of off diagonals). 
Thus the latter plays a similar role as logarithmic negativity plays in entanglement theory 
and this is the best operational interpretation from a resource-theoretic viewpoint. 
Consequently the two measures are intimately connected to another operational measure, 
the robustness of coherence. We find also relationships between these measures, which are tight for general states, and the tightest possible for pure and qubit states. For a given robustness, we construct a state having minimum distillable coherence. 
\end{abstract}


\maketitle

\section{Introduction}
 A single quantum system, where the notion of entanglement is meaningless, nonetheless, differs in many ways from a classical system. Indeed, the inception of quantum mechanics itself was triggered by phenomena such as interference, wave-particle duality, observed on single system. In such cases, the figure of merit is attributed to the superposition principle, i.e., the characteristic of quantum mechanics which allows superposition of prefixed basis states as a valid state. This coherence, one of the fundamental reasons for many counter-intuitive features of quantum mechanics, allows precise description at mesoscopic scales. In general, coherence is an important physical resource in single-particle interferometry \cite{Oi.PRL.2003,Aberg.PRA.2004,Oi+Aberg.PRL.2006}, quantum thermodynamics \cite{Skrzypczyk+2.NC.2014,Aberg.PRL.2014,Lostaglio+2.NC.2015,Narasimhachar+Gour.NC.2015,Lostaglio+3.PRX.2015,Cwiklinski+3.PRL.2015,Korzekwa+3.NJP.2016}, spin systems \cite{Karpat+2.PRB.2014,Malvezzi+5.PRB.2016}, nanoscience \cite{Vazquez+4.NN.2012,Braakman+4.NN.2013,Caram+8.JPCL.2014}, quantum algorithms \cite{Hillery.PRA.2016,Ma+4.PRL.2016,Matera+3.QST.2016}, and even some biomolecular processes \cite{Plenio+Huelga.NJP.2008,Rebentrost+2.JPCB.2009,Lloyd.JPCS.2011,Li+4.SR.2012,Huelga+Plenio.CP.2013}.  With such an ample usefulness, it is desirable to have a modern resource-theoretic approach to coherence. Recently one such framework has been put forward \cite{Aberg.Ar.2006,Baumgratz+2.PRL.2014}, which has been subsequently developed \cite{Winter+Yang.PRL.2016} and advanced further \cite{Streltsov+4.PRL.2015,Chitambar+5.PRL.2016,Streltsov+3.PRX.2017,Bromley+2.PRL.2015,Chitambar+Hsieh.PRL.2016,Ma+4.PRL.2016,Yadin+4.PRX.2016,Streltsov+4.Ar.2016}. For many other models, applications, and further details of coherence theory see the review in Ref.~\cite{Streltsov+2.RMP.2017}.

Undoubtedly, the monotones are an important aspect of any resource theory. 
On one hand, they certify impossibility of converting resources,
while on the other, they induce a partial order among the resource states. 
In the framework proposed in Ref.~\cite{Baumgratz+2.PRL.2014}, among the most 
interesting coherence monotones are the  $\ell_1$-norm-based coherence ($C_{\ell_1}$) \cite{Baumgratz+2.PRL.2014}, 
the relative entropy of coherence ($C_r$) \cite{Baumgratz+2.PRL.2014}, and the robustness of
coherence ($C_R$) \cite{Napoli+5.PRL.2016,Piani+5.PRA.2016}, which are formally defined as
follows:
\begin{align*}
 C_{\ell_1}(\rho )&:= \sum_{i\neq j} |\rho_{i,j}|, \\
 C_r(\rho)        &:= \min_{\delta\in\mathscr{I}} S(\rho\|\delta)
                    = S(\rho\|\diag(\rho)) = H(d)-H(\lambda) \label{Def:primary.quantities}\numberthis\\
 C_R(\rho)        &:= \min_{\sigma} \left\{s\geq 0 \,\Big|\, \frac{\rho+s\,\sigma}{1+s}\in\mathscr{I}\right\}
                    = \min_{\tau\in\mathscr{I}} \left\{s\geq 0 \,\big|\, \rho\leq(1+s)\tau\right\}
\end{align*}
where $\mathscr{I}$ is the set of incoherent states 
(i.e. states which are diagonal with respect to the chosen basis), 
$S(x\|y)=\tr[x(\log_2 x-\log_2 y)]$ is the relative entropy, 
$d$ and $\lambda$ are the vectors of diagonal elements and eigenvalues of $\rho$ 
respectively, and $H(p)=-\sum p_i\log_2(p_i)$ is the Shannon entropy of $p$. Both $C_r$ and $C_R$ have exact analogs in entanglement theory, 
both are operational quantities, and have direct physical significance. 
In contrast, $C_{\ell_1}$ is peculiar in the sense that it has no explicit prominent role in any other known resource theory so far (entanglement, nonlocality, discord, purity, etc), 
presumably due to its explicit dependence on the chosen basis. However, $C_{\ell_1}$ 
captures the simple intuitive idea that on the level of density matrix description 
of quantum mechanical states, superposition corresponds to off-diagonal matrix 
elements (always with respect to the selected basis). In fact, the $\ell_1$ norm has been used in a necessary condition for separability known as computable-cross-norm criterion \cite{Rudolph.PRA+LMP+QIP.2003}, and in quantification of a discord-like quantity named negativity of quantumness \cite{Nakano+2.PRA.2013}.
Physically, for instance, $C_{\ell_1}$ is responsible for the duality between 
fringe-visibility and which-path information in a two-path interferometer \cite{Bera+3.PRA.2015};
more generally, it also captures the which-path information about a particle inside 
a multipath interferometer \cite{Bagan+3.PRL.2016}. 

Motivated by this evident usefulness of $C_{\ell_1}$, in this work, 
we aim to give it an operational interpretation. 
Based on the facts that for pure states $C_{\ell_1}$ is equal (up to a factor of $2$) to negativity of the corresponding bipartite pure state, and both measures  satisfy strong monotonicity, we have surmised in \cite{Rana+2.PRA.2016} 
that $C_{\ell_1}$ is analogous to negativity in entanglement theory; we argued that if true, it would be one of the best operational interpretations 
of $C_{\ell_1}$. We show in this work that this is indeed the case. Our primary aim 
is thus to establish the sharpest possible interrelations between $C_r$ and $
C_{\ell_1}$. Keeping this in mind, we develop our results in steps, 
starting from the simplest qubit case, then pure states, and finally
general states. Conditions for equality as well as interrelations with other 
operational monotones (mainly $C_R$) will be mentioned along the way.  

\section{Qubit case}
Using an inequality between Holevo information and trace norm, it was shown in 
Ref.~\cite{Rana+2.PRA.2016} that all qubit states satisfy $C_r(\rho)\leq C_{\ell_1}(\rho)$. 
Several proofs of this fact will be given throughout this article.  
However, this is not the sharpest possible interrelation, as strict inequality 
occurs for almost all states. The following result represents the sharpest interrelations. 
\begin{proposition}
  \label{Prop:all.bound.for.qubits} 
  All qubit states  $\rho$  with a given coherence $C_{\ell_1}(\rho)=2b$ satisfy
  \begin{equation}
    \label{Eq:d=2.all.bounds} 
    1-H_2\left(\frac{1-2b}{2}\right) \leq C_r(\rho) \leq H_2\left(\frac{1-\sqrt{1-4b^2}}{2}\right) 
    \leq C_{\ell_1}(\rho),
   \end{equation}
  where $H_2(0\le x\le 1):=-x\log_2(x)-(1-x)\log_2(1-x)$ is the binary entropy function.The lower and upper bounds on $C_r$ are saturated by a unique state for each $b$
  (up to incoherent unitaries). Equality holds in all the inequalities iff $\rho$ is either an incoherent or a 
  maximally coherent state, otherwise $C_r(\rho)<C_{\ell_1}(\rho)$. 
\end{proposition}
 The proof uses convexity of $C_r$ and is given in 
 Appendix~\ref{App:Proof:Prop:all.bound.for.qubits}. It also uses the following well-known inequality 
for binary entropy:
\begin{equation}
	\label{Eq:Bounds.on.H2}2\min\{x,1-x\}\leq H_2(x)\leq 2\sqrt{x(1-x)},\quad \forall~x\in[0,1].
\end{equation} But this inequality alone does not yield even the crude bound $C_r(\rho)\leq C_{\ell_1}(\rho)$.

\section{Pure states}
We showed in \cite{Rana+2.PRA.2016} that all pure states also satisfy 
$C_r\leq C_{\ell_1}$. An independent proof was also given in \cite{Chen+4.PRA.2016}. 
We first characterize the equality conditions. 
\begin{proposition}
  \label{Prop:Cr=Cl1.iff} 
  All pure states satisfy $C_{\ell_1}(\rho)\geq C_r(\rho)$. 
  Equality holds iff the diagonal elements are (up to permutation) 
  either $\{1,0,\cdots,0\}$, or, $\{1/2,1/2,0,\dotsc,0\}$.
\end{proposition}
Proof: Using the recursive property \cite{Lin.IEEE.1991} of entropy function $H(\lambda)$ , we have
\begin{align*}
&C_{\ell_1}\left(|\psi\ran=\sum_{i=1}^d\sqrt{\lambda_i}|i\ran\right)-C_r(|\psi\ran)\\
&=2\sum_{i=1}^{d-1}\sqrt{\lambda_i}\sum_{j=i+1}^d\sqrt{\lambda_j}-H(\lambda)\\
&\geq 2\sum_{i=1}^{d-1}\sqrt{\lambda_i}\sqrt{\sum_{j=i+1}^d\lambda_j}-H(\lambda)\label{Eq:sqrt.x.is.concave}\numberthis\\
&=\sum_{i=1}^{d-1}\left[\left(\sum_{k=i}^d\lambda_k\right)\left(2\sqrt{\frac{\lambda_i}{\sum_{k=i}^d\lambda_k}\left(1-\frac{\lambda_i}{\sum_{k=i}^d\lambda_k}\right)}-H_2\left(\frac{\lambda_i}{\sum_{k=i}^d\lambda_k}\right)\right)\right].
\end{align*}
By inequality \eqref{Eq:Bounds.on.H2}, each term in the above sum is non-negative, so for vanishing of the sum, each term should vanish. For equality in Eq.~\eqref{Eq:sqrt.x.is.concave}, only two of the $\lambda_i$'s could be nonzero. Vanishing of the first term yields $\lambda_1=1,0,1/2$. \qed

Now we give an upper bound for the difference $C_{\ell_1}-C_r$ and present an alternative proof, arguably the simplest one, for its lower bound.
\begin{proposition}
  \label{Prop:Max.of.Cl1-cr.pure} 
  For all pure states $|\psi\ran$ with $\rank[\diag(|\psi\ran\lan\psi|)]=d>2$,
  \begin{equation}
    \label{Eq:Max.of.Cl1-cr.pure}
	0 \leq C_{\ell_1}(|\psi\ran)-C_r(|\psi\ran) \leq d-1-\log_2d.
  \end{equation}
\end{proposition}
The proof uses Schur-concavity of $ C_{\ell_1}(|\psi\ran)-C_r(|\psi\ran)$ in 
$\diag(|\psi\ran\lan|\psi|)$ and is detailed in Appendix~\ref{App:Proof:Max.of.Cl1-cr.pure}.

These bounds are rough in the sense that they do not require knowledge of either 
$C_r(\rho)$ or $C_{\ell_1}(\rho)$. A better bound follows below, whose
proof is given in  Appendix~\ref{App:Proof:Prop:Pure.Max.Cr.given.Cl1}.
\begin{proposition}
	\label{Prop:Pure.Max.Cr.given.Cl1} 
	For a given $\ell_1$-norm coherence $C_{\ell_1}(|\psi\ran)=b$, 
	$C_r(|\psi\ran)$ is bounded by
	\begin{equation}
	  \label{Bound:on.cr.given.cl1.}
	  \frac{\sqrt{2}b^2}{d(d-1)}\leq C_r\leq \log_2(1+b),
	\end{equation}
	where $d=\rank[\diag(|\psi\ran\lan\psi|)]$. The lower bound is saturated 
	only for incoherent states while the upper bound is saturated by incoherent 
	and maximally coherent states. 
\end{proposition}

The lower bound indicates that for a given $C_{\ell_1}$, the value of $C_r$  probably could be made arbitrarily small for high dimension. 
In contrast, the upper bound does not depend on the dimension $d$. Thus, for a fixed $C_{\ell_1}=b$, we can not increase $C_r$ beyond $\log_2(1+b)$ even  by increasing the dimension arbitrarily (but keeping it finite). 

This motivates the following question: what could be the sharpest (maximum and minimum) 
values of $C_r$ given only the knowledge of $C_{\ell_1}$? Fortunately, we are 
able to give the precise answer in the following.

\begin{theorem}
  \label{Thm:Tightest.bounds.pure} 
  All pure states $|\psi\ran$ with a given $C_{\ell_1}(|\psi\ran)=b$ satisfy
  \begin{align*}	
    H_2(\alpha) &     +(1-\alpha)\log_2(d-1)\leq C_r(|\psi\ran) \\
                &\leq H_2(\beta)+(1-\beta)\log_2(n-1),\label{Bound:tightest.possible.pure}\numberthis\\
    \text{where } 
    \alpha      &=    \frac{2+(d-2)(d-b)+2 \sqrt{(b+1) (d-1) (d-1-b)}}{d^2},\\
    \beta       &=    \frac{2+(n-2)(n-b)-2 \sqrt{(b+1) (n-1) (n-1-b)}}{n^2},\\
    d           &=    \rank[\diag(|\psi\ran\lan\psi|)],\\
    n           &=    \begin{cases}
                        b+1       & \quad \text{if } b \text{ is integer},\\
                        [b]+2  & \quad  \text{ otherwise},
                      \end{cases}
  \end{align*} 
  with $[x]$ denoting the integer part of $x$.

  Each of the bounds is satisfied by a unique state, up to permutation the 
  diagonal elements of the state with minimum $C_r$ are given by 
  $\{\alpha,(1-\alpha)/(d-1),(1-\alpha)/(d-1),\cdots,(1-\alpha)/(d-1)\}$ 
  and that with maximum $C_r$ are 
  $\{\beta,(1-\beta)/(n-1),(1-\beta)/(n-1),\cdots,(1-\beta)/(n-1)\}$. 
\end{theorem}
The proof is based on Lagrange multipliers, and uses some techniques recently 
employed in Refs.~\cite{Audenaert+2.JMP.2016,Correa+3.PRL.2005}. The complete proof is given in Appendix~\ref{App:Proof:min.}.  
In Fig.~\ref{Fig:fig1}, we show the several bounds on $C_r$ as a function of $C_{\ell_1}$. 

We note that for any fixed $b$, as $d\to\infty$ $\alpha\to 1$ and the lower bound of $C_r\to0$ . Thus, for any fixed value of $C_{\ell_1}=b$, there is a $|\psi\ran\in\CC^d$ (for sufficiently high $d$) with $C_{\ell_1}(|\psi\ran)=b$ and arbitrary small $C_r(|\psi\ran)$. In contrast, we can not increase $C_r$ beyond the upper bound (which depends on $b$ but is independent of dimension). An explanation is that given more and more components, the probability could be made more biased but not more uniform than the initial one.
\begin{figure}[h]
	\begin{center}
		\includegraphics[height=5cm]{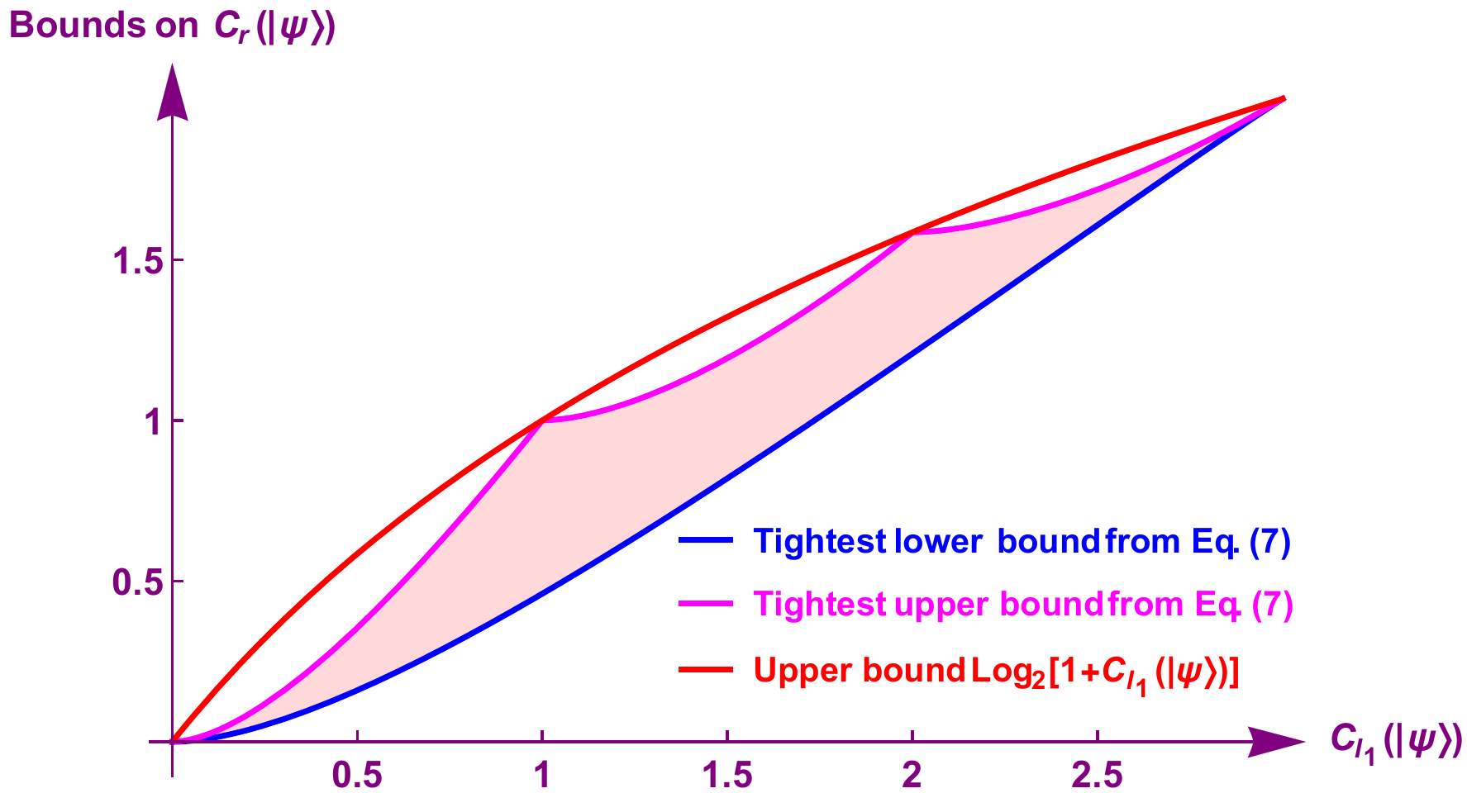}\caption{(Color online) $C_r(|\psi\ran)$ vs. $C_{\ell_1}(|\psi\ran)$ for (normalized) $|\psi\ran\in\CC^4$: given only $C_{\ell_1}$ and $d$, the bounds in Eq.~\eqref{Bound:tightest.possible.pure} are the tightest possible. For any point $(x,y)$ inside the pink region (including the boundary curves), there is a $|\psi\ran$ such that $x=C_{\ell_1}(|\psi\ran)$ and $y=C_r(|\psi\ran)$.} 
		\label{Fig:fig1}
	\end{center}
\end{figure}

This result for pure states has an interesting aspect: since $C_{\ell_1}=C_R$ for all 
pure states \cite[Theorem~6]{Piani+5.PRA.2016}, Theorem~\ref{Thm:Tightest.bounds.pure} 
also gives the sharpest bounds on $C_r$, for a given robustness $C_R=b$. Note also that unless $C_{\ell_1}$ has an integral value, no pure state saturates the inequality $C_r\leq \log_2(1+C_{\ell_1})$.

\section{Arbitrary states} 
As usual, the case of mixed states is more demanding, since in this case $C_r$ 
depends on the eigenvalues, which are implicit functions of the matrix elements. 
Another difficulty is that the quantity $C_{\ell_1}$ is not unitarily invariant. 
So, we have to resort to different techniques. 
But, before dealing with general mixed states, let us mention that the result 
$C_r\leq C_{\ell_1}$ holds for the following simple class of states. 
\begin{proposition}
	\label{Prop:Case.of.pseudo.mixture} Any pseudopure state of the form 
	$\rho=p|\psi\ran\lan\psi|+(1-p)\delta$ with $p\in[0,1]$ and $\delta\in\mathscr{I}$ 
	satisfies $C_r(\rho)\leq C_{\ell_1}(\rho)$. 
	This gives an alternative proof for the validity of the same relation for any qubit-state.
\end{proposition}
\noindent Proof: From the convexity of $C_r$, we have
\begin{align*}
C_r(\rho)&\leq pC_r(|\psi\ran\lan\psi|)\leq pC_{\ell_1}(|\psi\ran\lan\psi|)\\ 
         &=C_{\ell_1}(p|\psi\ran\lan\psi|+(1-p)\delta)=C_{\ell_1}(\rho).
\end{align*}
Since every (mixed) qubit state can be expressed as the above pseudomixture, the result follows.\qed

We have seen that for pure states when $C_{\ell_1}>1$,  $C_{\ell_1}$ is too high compared to $C_r$ and so $\log_2(1+C_{\ell_1})$ is reasonably a better upper bound for $C_r$.  It turns out that the same upper bound holds also for all (mixed) states.

\begin{theorem}
	\label{Thm:Gen.Upper.mixed} 
	For any state $\rho$,
	\begin{equation}
	  \label{Eq:Ge.upper.mixed}
	  C_r(\rho) \leq \log_2\big[1+C_R(\rho)\big] \leq \log_2\big[1+C_{\ell_1}(\rho)\big].
	\end{equation}
	\end{theorem}
The proof uses operator monotonicity of $\log$ function and the details are presented in Appendix~\ref{App:Proof:Ge.upper.mixed}. Here we give an alternative proof for $C_r\leq\log_2[1+C_{\ell_1}]$, highlighting the similarity of $C_{\ell_1}$ with negativity.
Recalling that distillable entanglement $E_d$ is upper bounded 
by the logarithmic negativity \cite{Vidal+Werner.PRA.2002}, and for any state, 
coherent information is upper bounded by one-way distillable entanglement $E_{\to}$ (by
the so-called \emph{hashing inequality}, \cite[Theorem 10]{Devetak+Winter.PRSLA.2005}), 
we get 
\begin{equation}
  \label{Eq:Hashing.inequality.and.Negativity}
  S(\sigma^A)-S(\sigma^{AB})\leq E_{\to}(\sigma)\leq E_d(\sigma)\leq\log_2\big[1+2\N(\sigma)\big].
\end{equation}
For any given $\rho=\sum a_{ij}|i\ran\lan j|$, consider the state $\sigma^{AB}=\sum a_{ij}|ii\ran\lan jj|$. One immediately verifies that $S(\sigma^A)-S(\sigma^{AB})=C_r(\rho)$. The eigenvalues of partial transposition of $\sigma^{AB}$ are $a_{ii}$ for $i=1,2,\cdots,d$, and $\pm|a_{ij}|$ for $1\leq i<j\leq d$ \cite[Lemma~6.3]{Rains.IEEE.2001}. Therefore, $2\N(\sigma)=2\sum_{i<j}|a_{ij}|=C_{\ell_1}(\rho)$. Substituting these values in Eq.~\eqref{Eq:Hashing.inequality.and.Negativity}, we get the desired result.\qed

Yet another method to prove the same inequality is to use the monotonicity of sandwiched $\alpha$-R\'{e}nyi relative entropy
\[S_{\alpha}(A\|B):=\frac{1}{\alpha-1}\log_2\tr \left[B^{\frac{1-\alpha}{2\alpha}}AB^{\frac{1-\alpha}{2\alpha}}\right]^\alpha\]
in $\alpha>0$ \cite{Hiai.Thanks}.

Since $\log_2(1+x)\leq x$ for all $x\geq 1$, we have by Eq.~\eqref{Eq:Ge.upper.mixed} \begin{equation}
\label{Eq:Cr.depending.on.Cl1}
C_r(\rho)\leq\left\{\begin{array}{ll}
C_{\ell_1}(\rho),&\text{ if } C_{\ell_1}(\rho)\geq 1\\
C_{\ell_1}(\rho)\log_2e,&\text{ if } C_{\ell_1}(\rho)<1
\end{array}.\right.\end{equation}
Thus $C_r\leq C_{\ell_1}$ holds for all states, at most up to a multiplicative constant of $1/\ln2$. Unfortunately, we could not resolve the conjecture $C_r\leq C_{\ell_1}$ made in Ref.~\cite{Rana+2.PRA.2016} in full generality. However, employing perturbative techniques, we could prove it when $C_{\ell_1}$ is very small (see Appendix~\ref{App:Support.4.Conj}).
Note, on the other hand,  that if $C_r(\rho)\leq C_{\ell_1}(\rho)$ is true, then it is  the sharpest possible upper bound on $C_r$ when $C_{\ell_1}(\rho)\leq 1$.
\begin{proposition}
	\label{Prop:Max.Cr=Cl1=b<1} 
	For any $0<b<1$ and $d\geq 3$, there is a $d$-dimensional state 
	$\rho$ with $C_r(\rho)=C_{\ell_1}(\rho)=b$.
\end{proposition}
\noindent Proof: One such state is given by 
\[
  \rho = \begin{pmatrix}
           b/2&b/2\\b/2&b/2
         \end{pmatrix}
         \oplus (1-b)\delta,
\]
with any $(d-2)$-dimensional diagonal state $\delta\in\mathscr{I}$.\qed

It is desirable to sharpen Eq.~\eqref{Eq:Ge.upper.mixed} to something like Eq.~\eqref{Bound:tightest.possible.pure}. However, we are not aware of any sharper bounds. Our numerical study suggests that, for a given $C_{\ell_1}$, the state with max $C_r$ is generally a mixed one, unless we put restriction also on the dimension (it is a pure state, if additionally $d\leq [C_{\ell_1}]+2$). 

Nonetheless, we have completely characterized the sharpest lower bound of $C_r$ for a given $C_R$. The next result guarantees the minimum amount of distillable coherence from a resource state given only the dimension $d$ and $C_R$.   
\begin{theorem}
	\label{Prop:Ultimate.bounds} All states $\rho$ with a given $C_R(\rho)=b$  satisfy
	\begin{align*}	
	C_r(\rho)&\geq \log_2d-H_2(\alpha)-(1-\alpha)\log_2(d-1)\label{Ultimate.bounds}\numberthis\\
	\text{where } d&= \rank[\diag(\rho)] \text{ and } 	\alpha=\frac{1+b}{d}.
	\end{align*}
	Equality occurs for isotropic-like states 
	$\rho=p|\Psi\ran\lan\Psi|+(1-p)\id/d$, $p=b/(d-1)$, and $|\Psi\ran$ 
	being the maximally coherent state.
\end{theorem}
The full proof is presented in Appendix~\ref{App:Prop:Ultimate.bounds}. Appendix~\ref{App:Failed.Attempt} contains 
an unsuccessful attempt to prove $C_r\leq C_{\ell_1}$ via convex roofs \cite{Qi+2.JPA.2017,Chin.PRA.2017}. Nevertheless, it could be of independent interest because of its close connection with the convex roof of negativity for maximally correlated states.

\section{Logarithmic coherence: a strong monotone which is not convex}
Similar to logarithmic negativity $E_{\N}$ \cite{Vidal+Werner.PRA.2002,Plenio.PRL.2005}, 
we can define                 
$C_{\log}(\rho):=\log_2[1+C_{\ell_1}(\rho)]$. 
The addition by $1$ not only makes $C_{\log}\geq 0$, 
but also yields the additivity under tensor products,
$C_{\log}(\rho\otimes\sigma)=C_{\log}(\rho)+C_{\log}(\sigma)$, just like $C_r$ and $E_{\N}$. 
The strong monotonicity follows easily from that of $C_{\ell_1}$,
using concavity and monotonicity of the logarithm: 
%If $p_i$s are probabilities and $a_i\geq0$, the weighted arithmetic-geometric 
%inequality reads $\prod a_i^{p_i}\leq\sum p_ia_i$. Thus,
\begin{align*}
  \sum_i p_i \log_2 \bigl[1+C_{\ell_1}(\rho_i)\bigr]
    &\leq \log_2 \left[\sum_i p_i \bigl[1+C_{\ell_1}(\rho_i)\bigr] \right] \\
    &=    \log_2 \left[1 + \sum_i p_i C_{\ell_1}(\rho_i) \right] \\
    &\leq \log_2\bigl[1+C_{\ell_1}(\rho)\bigr],
\end{align*}
the last inequality due to strong monotonicity of $C_{\ell_1}$. 
%Since $\log$ is a monotonically increasing function, applying $\log_2$ both sides gives the strong monotonicity of $C_{\log}$. 
Due to the concavity of $\log$ function, however, $C_{\log}$ is not convex:
\[
  C_{\log}\left(\frac{1}{2}\rho+\frac{1}{2}\sigma\right)
    > \frac{1}{2}C_{\log}(\rho)+\frac{1}{2}C_{\log}(\sigma),
\]
iff $C_{\ell_1}(\rho)C_{\ell_1}(\sigma)[C_{\ell_1}(\rho)- C_{\ell_1}(\sigma)]\neq 0$.
Note that the above arguments show that for any strong monotone $C$, 
the logarithmic version $C_{\log}=\log_2(1+C)$ is also a (nonconvex) 
strong monotone; there is nothing special about $C_{\ell_1}$---except 
that in this case $C_{\log}$ is additive under tensor products. 

\section{Relevance}
The main importance of this work is that it gives operational interpretation to 
$C_{\ell_1}$ in a completely quantitative way, namely it is similar to negativity 
in entanglement theory, and indeed the logarithmic coherence defined here, 
though not convex, is a better motivated one. The latter plays the exact role of logarithmic negativity in entanglement theory, giving a tight upper bound on distillable resource. Once this is established, we can seamlessly browse all instances of usefulness of  (logarithmic) negativity as an entanglement monotone from entanglement theory to coherence theory. For example, Theorem~\ref{Thm:Gen.Upper.mixed} is just a manifestation of known interrelations between relative entropy of entanglement, (logarithmic) negativity, and robustness of entanglement. Thus $C_{\ell_1}$, though arguably one of the simplest monotones which has apparently no conspicuous role in entanglement theory, is significant for most relevant operational quantities in coherence theory. Later we will mention relevance of our results beyond a particular resource theory. 

In many practical scenarios, the density matrix depends on some parameters (e.g., the entries are functions of time---in time-dependent evolution; temperature or other relevant parameters---in thermometry or metrology). In such cases, the density matrix cannot be diagonalized and hence $C_r$ becomes uncomputable. The precise bounds given in this work are the best from the knowledge of the entries.  

We would also like to mention possible applications of our results to some related fields, namely information theory and matrix analysis. First note that $C_{\ell_1}(|\psi\ran)$ is the R\'{e}nyi entropy 
\[
  R_{\alpha}(\lambda) := \frac{1}{1-\alpha}\log_2\left[\sum_{i=1}^{d}\lambda_i^{\alpha}\right]
                       = \frac{\alpha}{1-\alpha}\log_2\big(\norm{\lambda}_{\alpha}\big), 
\]
of order $\alpha=1/2$ in disguise. Thus the relation between $C_r(|\psi\ran)$ and 
$C_{\ell_1}(|\psi\ran)$ is actually optimizing $R_\alpha$ (${\alpha\to 1}$) 
subject to the given fixed value of $R_{1/2}$. The upper bound in 
Eq.~\eqref{Bound:on.cr.given.cl1.} is just a consequence of nonincreasing 
property of $R_{\alpha}$. The optimization technique employed in Appendix~\ref{App:Proof:min.} could also be applicable to other values of $\alpha$. 
Indeed, it is easy to find sharpest bounds on $R_{\alpha\to 1}$ subjected to a 
fixed $R_{2}$, which reproduce the result from \cite{Harremoes+Topsoe.IEEE.2001}.  

Lastly, finding trade-off relations between diagonals, eigenvalues, singular values, etc, 
are standard problems in matrix analysis \cite[Ch.~9]{Marshall+2.Book.S.2011}. 
Our main quest here was a small part, finding exact trade-off between diagonals 
and eigenvalues (via entropy function), having the knowledge of the sum of absolute 
values of the entries. One such independent relation is Eq.~\eqref{Ultimate.bounds}
(it is worth mentioning that the same matrix maximizes the determinant \cite{Neubauer.LAA.1997}, 
a log-concave function). More precisely, our problem is exactly similar to 
finding sharpest Fannes-Audenaert bound \cite{Fannes.CMP.1973,Audenaert.JPA.2007} 
for a single state and our results are independent of similar 
bounds \cite{Audenaert+Eisert.JMP.2005,Audenaert.JMP.2014}. 

\section{Discussion and conclusion}
We have shown that in the coherence theory \cite{Baumgratz+2.PRL.2014}, 
$C_{\ell_1}$ operationally plays the exact role of negativity in entanglement theory.  
Since there is no bound coherence \cite{Winter+Yang.PRL.2016} (analogous to no bound entanglement 
in maximally correlated states), $C_{\ell_1}$ is intimately connected to any 
operationally relevant quantity or process. For example, the sharpest bounds on 
$C_r$ from Theorem~\ref{Thm:Tightest.bounds.pure} remain the same even if we 
replace $C_{\ell_1}$ by $C_R$. Thus, our approach here supports the idea 
that coherence theory is a subclass of entanglement theory for maximally 
correlated states. Nonetheless, similar to entanglement theory, we showed that
the requirement of convexity, although a desirable property, should be relaxed for coherence monotones.

Given their similar operational meaning, it would be interesting to compare Eq.~\eqref{Thm:Gen.Upper.mixed} with its entanglement-analog $E_d=\log_2[1+2\N]$, especially since in contrast to $C_d=C_r$, $E_d$ is a noncomputable quantity. Note that for the NPT bound entangled states [proof of whose (non-)existence is an open problem in quantum information theory, with all conjectures in literature claiming the existence \cite{OPiQIT.page}], the bound on $E_d$ is worst as it gives absolutely no information. However, the relation for $C_d$ always gives some nonzero bound, thereby the inequality has more to offer in coherence theory. From quantitative perspectives, both bounds are quite rough as almost all the states never achieve equality. Our results in Theorem~\ref{Thm:Tightest.bounds.pure} and Theorem~\ref{Prop:Ultimate.bounds} are the best possible in this regard, as they give the optimal bound on one quantity from the knowledge of the other.
 
It is worth remarking that the relation $C_{\ell_1}\geq C_r$ does not hold for normalized quantities. The normalized quantities, being dimension dependent, need not be monotone. Also, $C_r$ and $C_{\ell_1}$ do not give the same ordering of state space. For example, there are states $\rho$ and $\sigma$ such that $C_{\ell_1}(\rho)>C_{\ell_1}(\sigma)>C_r(\sigma)>C_r(\rho)$.

Before concluding, we would like to mention that over the past two years many alternative frameworks of coherence theory have been proposed \cite{Chitambar+Gour.PRL.2016,Marvian+Spekkens.PRA.2016,Yadin+Vedral.PRA.2016,Vicente+Streltsov.JPA.2016,Yu+3.AR.2016}, stemming mainly from different notions of incoherent (free) operations. In some of these models, $C_{\ell_1}$ is not a monotone and arguably there is no maximally coherent state \cite[Table~II]{Streltsov+2.RMP.2017}, thereby lacking the interpretation of $C_r$ as distillable coherence. However, both $C_r$ and $C_R$ are not only monotones, but also operational quantities  even in the most general (reversible) resource theory \cite{Brandao+Plenio.CMP.2010,Brandao+Gour.PRL.2015}. Most of our results, as could also be seen as relations between $C_r$ and $C_R$, are thus applicable to more general scenarios. Pertinent to coherence, the most general framework by \AA{}berg \cite{Aberg.Ar.2006}, where incoherent states are block-diagonal of any block size, allows an interrelation analogous to Eq.~\eqref{Eq:Ge.upper.mixed}; we have to replace $C_{\ell_1}$ by the sum of trace norm of all off-diagonal blocks \cite{Hiai.Thanks}.

\section*{Acknowledgments}
We thank Tsuyoshi Ando and Fumio Hiai for many helpful correspondences and Nilanjana Datta, Alexander Streltsov, Ludovico Lami, and Manabendra Nath Bera for inspiring discussions. We also thank Nathaniel Johnston for bringing Ref.~\cite{Rudolph.PRA+LMP+QIP.2003} to our attention.

SR and ML acknowledge financial support from the John Templeton Foundation, 
the European Commission grants OSYRIS (ERC-2013-AdG Grant No. 339106), 
QUIC (H2020-FETPROACT-2014 No. 641122), and SIQS (FP7-ICT-2011-9 No. 600645), 
the Spanish MINECO grant FISICATEAMO (FIS2016-79508-P), the ``Severo Ochoa'' Programme 
(SEV-2015-0522), and MINECO\, CLUSTER (ICFO15-EE-3785), the Generalitat de Catalunya 
(2014 SGR 874 and CERCA/Program), and Fundaci\'{o} Privada Cellex. 
AW is supported by the European Research Council (AdG IRQUAT No. 267386), the European 
Commission (STREP RAQUEL FP7-ICT-2013-C-323970), the Spanish MINECO (Project 
No.~FIS2013-40627-P), and the Generalitat de Catalunya (CIRIT Project No.~2014 SGR 966). 

\noindent\emph{Note added}: The operational interpretation of $C_{\ell_1}$ presented in this work has been complemented in Ref.~\cite{Zhu+2.AR.2016}.

\newpage

\appendix

\section{Proof of Proposition~\ref{Prop:all.bound.for.qubits}}
\label{App:Proof:Prop:all.bound.for.qubits}

\begin{figure}[h]
	\begin{center}
		\includegraphics[height=4.5cm]{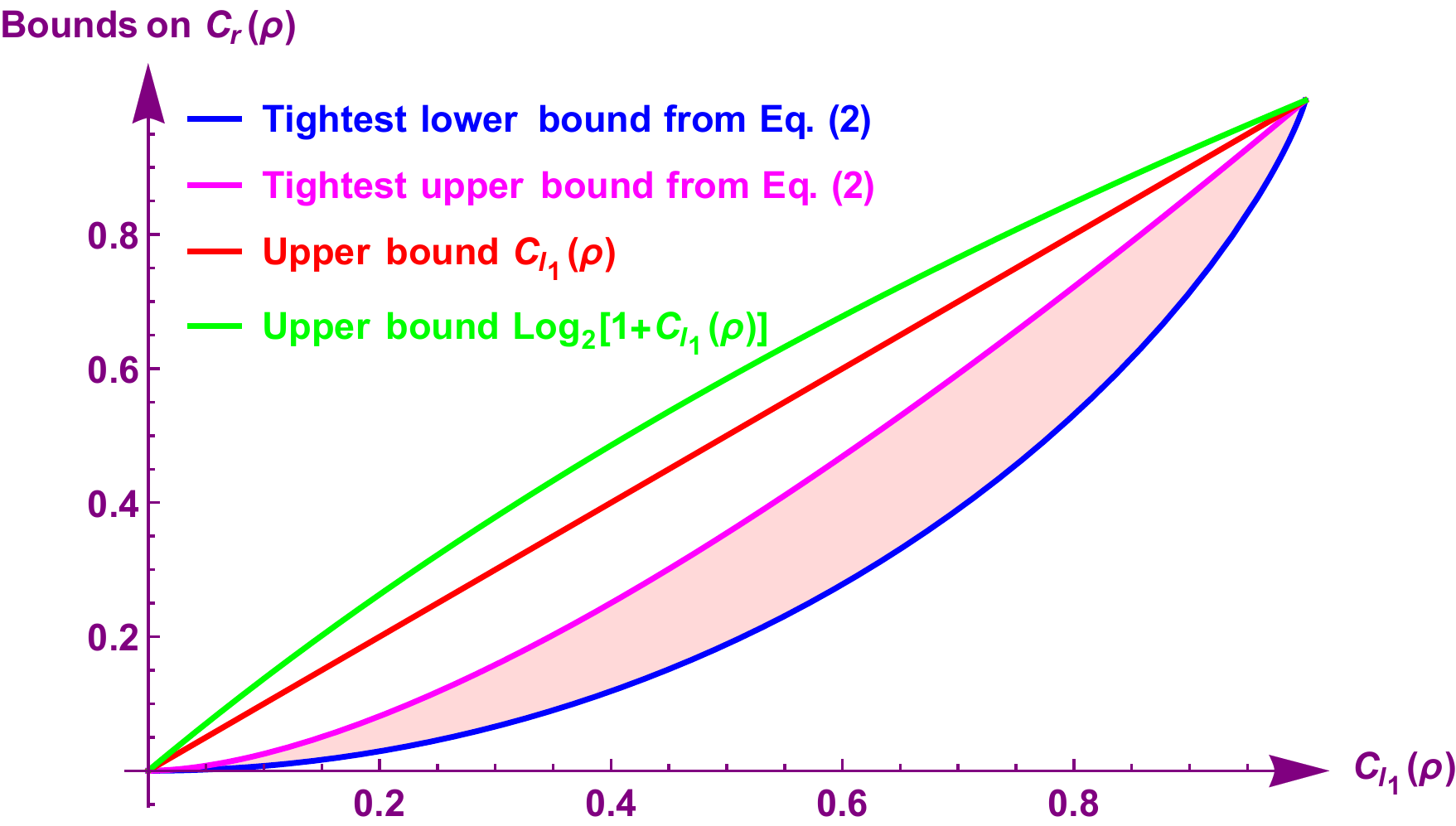}
		\caption{(Color online) $C_r(\rho)$ vs. $C_{\ell_1}(\rho)$ for qubit $\rho$: the bounds given by Eq.~\eqref{Eq:d=2.all.bounds} are the tightest possible. For any point $(x,y)$ inside the pink region (or over the boundary curves), there is a qubit state $\rho$ such that $x=C_{\ell_1}(\rho)$ and $y=C_r(\rho)$. Note that for a given $C_{\ell_1}$, there is a unique pure state, whose $C_r$ is given by the tightest upper bound (the magenta colored curve).} 
		\label{Fig:fig2}
	\end{center}
\end{figure}

\begin{proof}
	Without loss of generality, let 
	\[
	\rho=\begin{pmatrix}
	a&b\\b&1-a
	\end{pmatrix}
	\]
	be a state with given $\ell_1$-norm coherence $2b>0$. 
	For positivity of $\rho$, we must have \begin{equation}
		\label{Eq:Cond.for.a.in.terms.of.b}\frac{1-\sqrt{1-4b^2}}{2}\leq a\leq \frac{1+\sqrt{1-4b^2}}{2},~0<b\leq\frac{1}{2}.
	\end{equation}
	We will now show that for fixed $b$, $C_r(\rho):=H_2(a)-H_2(\lambda)$ is a convex function of $a$ over the entire region \eqref{Eq:Cond.for.a.in.terms.of.b}.
	To this end, the double derivative of $C_r$ with respect to $a$ is given by
	\begin{widetext}
		\[\frac{b^2 \left[\sqrt{1-4(a(1-a)-b^2)}\left(8 a^2-8 a+4 b^2+1\right)+8a (1-a)\left(a (1-a)-b^2\right) \ln \left(\frac{1+\sqrt{1-4(a(1-a)-b^2)}}{1-\sqrt{1-4(a(1-a)-b^2)}}\right)\right]}{a(1-a)\left[a(1-a)-b^2\right] \left[1-4(a(1-a)-b^2)\right]^{3/2}\ln2}.\]
	\end{widetext}
	Applying the inequality $\ln[(1+x)/(1-x)]\geq 2x$ for $x\in[0,1]$, the numerator is bounded below by
	$ b^2(1-2 a)^2 [1-4(a(1-a)-b^2)]^{3/2}$, a non-negative  quantity. Therefore $C_r(\rho)$ is convex and hence the maximum value will be attained at the extreme values of $a$ (and the corresponding state is a pure state). Thus for a given fixed $b$,
	we have 
	\begin{equation}
		\label{Eq:2qubit.max.cr.given.cl} C_r(\rho)\leq H_2\left(\frac{1-\sqrt{1-4b^2}}{2}\right).
	\end{equation}
	The upper bound on $H_2(x)$ from Eq.~\eqref{Eq:Bounds.on.H2} gives the right most inequality of Eq.~\eqref{Eq:d=2.all.bounds},
	\begin{equation}
		\label{Eq:2qubit.cr.leq.cl1} C_r(\rho)\leq H_2\left(\frac{1-\sqrt{1-4b^2}}{2}\right)\leq 2b=C_{\ell_1}(\rho).
	\end{equation}
	Note that for any given $b$ there is a $\rho$ (indeed a pure state) such that equality occurs in the first inequality of Eq.~\eqref{Eq:2qubit.cr.leq.cl1}, while except for incoherent states and maximally coherent states, the last inequality is always strict.
	
	The expression of $C_r$ remains unchanged if we interchange $a$ and $(1-a)$, i.e., $C_r$ is symmetric about $a=1/2$. Also, from Eq.~\eqref{Eq:Cond.for.a.in.terms.of.b} the allowed range of $a$ is symmetric about $a=1/2$. Therefore, $C_r$ being a symmetric convex function would have a unique global minimum at $a=1/2$. Hence the first inequality in Eq.~\eqref{Eq:d=2.all.bounds}. All the bounds for qubit systems are depicted in
	Fig.~\ref{Fig:fig2}.
\end{proof}

\section{Proof of Proposition~\ref{Prop:Max.of.Cl1-cr.pure}}
\label{App:Proof:Max.of.Cl1-cr.pure}

\begin{proof}
	Without loss of generality, let $|\psi\ran:=\sum_{i=1}^d\sqrt{\lambda_i}|i\ran$, 
	with $\lambda_i>0$ and $\sum\lambda_i=1$. We will now show that the function 
	\[
	f(\lambda):=C_{\ell_1}(|\psi\ran)-C_r(|\psi\ran)
	=\left(\sum_{i=1}^d\sqrt{\lambda_i}\right)^2-1+\sum_{i=1}^d\lambda_i\log_2\lambda_i
	\] is Schur-concave in $\lambda$, which will complete the proof.
	One verifies that 
	\begin{align*}
		\frac{\partial f}{\partial \lambda_1}-\frac{\partial f}{\partial \lambda_2}&=\frac{(\sqrt{\lambda_2}-\sqrt{\lambda_1})}{\sqrt{\lambda_1\lambda_2}}\left(\sum_{i=1}^d\sqrt{\lambda_i}\right)+\log_2\left(\frac{\lambda_1}{\lambda_2}\right)
		\\ &=(\sqrt{\lambda_2}-\sqrt{\lambda_1})\left[\frac{\sum_{i=1}^d\sqrt{\lambda_i}}{\sqrt{\lambda_1\lambda_2}}
		-\frac{\log_2\left(\sqrt{\frac{\lambda_1}{\lambda_2}}\right)}{2(\sqrt{\lambda_1}-\sqrt{\lambda_2})}\right].
	\end{align*}
	Thus, it suffices to show that the quantity inside the brackets is non-negative. 
	Using the \emph{geometric-logarithmic-mean inequality} \cite[p.~141]{Marshall+2.Book.S.2011}, 
	we get
	\[
	-\frac{\log_2\left(\sqrt{\frac{\lambda_1}{\lambda_2}}\right)}{2(\sqrt{\lambda_1}-\sqrt{\lambda_2})}
	\geq-\frac{1}{2\ln2\,(\lambda_1\lambda_2)^{1/4}}>-\frac{1}{(\lambda_1\lambda_2)^{1/4}},
	\] 
	and hence the quantity inside the brackets is non-negative.
\end{proof}
A sufficient (but not necessary) condition for a Schur-concave function $\phi$ to satisfy $\phi(x)>\phi(y)$ whenever $x\preceq y$ and $y$ is not a permutation of $x$ is that $\phi$ is \emph{strictly} Schur-concave \cite[p.~83]{Marshall+2.Book.S.2011}. Although both $C_{\ell_1}(\lambda)$ and $C_r(\lambda)$ are strictly Schur-concave, $f(\lambda)$ is not. This makes it difficult to characterize the equality conditions in Eq.~\eqref{Eq:Max.of.Cl1-cr.pure}. Nonetheless, saturation of the lower bound has been fully characterized in Proposition~\ref{Prop:Cr=Cl1.iff}. It is tempting to think that the upper bound will be saturated only by maximally coherent states if $d>2$. Although it could be true for $d\geq 4$, there are many $\lambda$'s giving the same maximum of $f$, with  $\lambda=(2/3,1/6,1/6)$ being an example for $d=3$.

\section{\label{App:Proof:Prop:Pure.Max.Cr.given.Cl1}Proof of Proposition~\ref{Prop:Pure.Max.Cr.given.Cl1}}

Using the inequality \begin{equation}
	\label{Eq:Bound:on.x.log2.x}-x\log_2x\geq\sqrt{2}x(1-x)\quad \,\forall x\in[0,1]
\end{equation}  we get
\begin{align*}
	C_r(|\psi\ran)&=\sum_{i=1}^d-\lambda_i\log_2\lambda_i\\
	&\geq\sqrt{2}\left[1-\sum_{i=1}^d\lambda_i^2\right]=\sqrt{2}\sum_{i\neq j}\lambda_i\lambda_j\\
	&\geq\frac{\sqrt{2}}{d(d-1)}\left(\sum_{i\neq j}\sqrt{\lambda_i\lambda_j}\right)^2\\
	&=\frac{\sqrt{2}b^2}{d(d-1)},
\end{align*}
where in the last inequality, we have used the fact that for a $d$-dimensional vector $x$, and for $0<p<q<\infty$, $\norm{x}_p\leq d^{1/p-1/q}\norm{x}_q$.

One weakness of this bound is that equality holds for incoherent states only. A lower bound on $C_r$, which is saturated by all incoherent and maximally coherent states, can also be derived easily. For example, using the following bound on entropy \cite{Dragomir+Goh.MCM.1996},  \[H(\lambda)\geq\log_2d-\frac{1}{\ln 2}\left[d\left(\sum_{i=1}^d\lambda_i^2\right)-1\right],\] we get \begin{equation}
	\label{Eq:bound:on.cr.pure.lower} C_r\geq\log_2 d-\frac{(d-1)^2-b^2}{(d-1)\ln 2}.
\end{equation}
Note that this lower bound is useful only when $b>\sqrt{(d-1)[(d-1)-\ln d]}$.

Now, to get the upper bound, we use concavity of logarithm,
\begin{align*}
	C_r(|\psi\ran)=H(\lambda)&=2\sum_{i=1}^d\lambda_i\log_2(1/\sqrt{\lambda_i})\\
	&\leq \log_2\left[\left(\sum_{i=1}^d\sqrt{\lambda_i}\right)^2\right]\\
	&=\log_2(1+b).
\end{align*}

To prove inequality \eqref{Eq:Bound:on.x.log2.x}, note that for $x\in(0,1)$, $\ln x=\ln[1-(1-x)]=-(1-x)-(1-x)^2/2-\ldots\leq -(1-x)$. Multiplying by $x/\ln 2$ and noticing that $1/\ln 2>\sqrt{2}$, the inequality follows.\qed

\section{\label{App:Proof:min.} Proof of Theorem~\ref{Thm:Tightest.bounds.pure}}
To prove the bounds we will optimize the entropy function w.r.t. the two equality constraints. The objective function being continuous, bounded (over the probability simplex $\Delta_d$, for a given dimension $d$), and the constraints describing compact sets, there is a maximum and a minimum. The optimum points should be either at interior or at the boundary of $\Delta_d$. As $\lambda_i=0$ neither affects the constraint, nor the objective function, if the optimum occurs on the boundary of $\Delta_d$, it should occur in the interior of $\Delta_n$ for some $n<d$. So, without loss of generality, we can assume that the optimum occurs in the interior of some $\Delta_n$, and  use Lagrange's multiplier method to get the possible stationary points. For simplicity, we can consider the natural-logarithm-based entropy (as it is a constant multiple of the binary-based entropy) and the Lagrange's function is set to be
\[\mathcal{L}(\lambda,\mu,\nu):=-\sum_{i=1}^n\lambda_i\ln\lambda_i+\mu\left[\sum_{i=1}^n\sqrt{\lambda_i}-\sqrt{1+b}\right]+\nu\left[\sum_{i=1}^n\lambda_i-1\right].\] 
Vanishing of the gradient ($\nabla\mathcal{L}=0$) gives
\begin{equation}\label{Eq:Lagrange.multipliers.min}
	1+\ln\lambda_i-\frac{\mu}{2\sqrt{\lambda_i}}-\nu=0.
\end{equation}

Solving Eq.~\eqref{Eq:Lagrange.multipliers.min} analytically is difficult. Instead, let us show that when seen as an equation in a particular $\lambda_i\in(0,\,1)$, it can have at most two (non-negative) solutions. The equation can be written as
\[z\ln z=a,\quad z=\sqrt{\lambda_i e^{1-\nu}},\,a=\frac{\mu\sqrt{e^{1-\nu}}}{4}.\]
The function $z\ln z$ is strictly convex in $(0,\infty)$ with a unique global minimum at $z=1/e$. So for any given $a>0$, the equation $z\ln z=a$ has a unique solution in $(1,\infty)$. However for any $a\in(-1/e,0)$ there are two solutions, one in $(0,1/e)$ and the other in $(1/e,1)$. Thus, overall there are at most two solutions to Eq.~\eqref{Eq:Lagrange.multipliers.min} for each $\lambda_i\in(0,1)$. Therefore, all possible stationary points of $\mathcal{L}$ must have two distinct values of $\lambda$'s, thereby potentially $k$-number of $\lambda_1$ and $(n-k)$-number of $\lambda_2$ with $0<\lambda_1<\lambda_2$, $k=1,2,\cdots,n$, $n\leq d$. Note that we could exclude the case $k=n$, as this is the case only when $b=n-1$, so there is only one pure state and no optimization is required. For a (given) finite $d$, this gives finite number of stationary points. However, as we will show below, we do not need to check all these points. Writing $x$ for $\lambda_1$, we get from normalization, $\lambda_2=(1-kx)/(n-k)$ and $x\leq 1/n$. Thus, the problem becomes 
\begin{align*}
	\text{Optimize }\quad   f(x,k)&:=-kx\ln x - (1-kx)\ln\left[\frac{1-kx}{n-k}\right]\\
	\text{s.t.     }\quad g(x,k)&:=k\sqrt{x}+\sqrt{(n-k)(1-kx)}=\sqrt{1+b}
\end{align*}
over $0<x\leq 1/n$, $k\in\{1,2,\cdots,n-1\}$, $n\leq d$. We will now employ the approach from \cite[Lemma~15]{Audenaert+2.JMP.2016}. 
To show that for a fixed $g(x,k)$, $f(x,k)$ is a decreasing function in $k$, let us temporally remove the integral restriction on $k$ and consider it as a real variable in $[0,n)$. Due to the constraint on $g(x,k)$, changing $k$ will also change $x$. So, to keep $g(x,k)$ fixed ($=\sqrt{1+b}$), let $x(k)$ be the function of $k$ implicitly given by $g(x(k),k)=\sqrt{1+b}$. Then $\frac{dg}{dk}=0=\frac{\pal g}{\pal k}+\frac{\pal g}{\pal x}.\frac{\pal x}{\pal k}$ gives
\[\frac{\pal x}{\pal k}=-\frac{\pal g}{\pal k}/\frac{\pal g}{\pal x}=\frac{x}{k}\left[\sqrt{\frac{1-kx}{(n-k)x}}-1\right].\]
Therefore, \begin{align*}
	\frac{df}{dk}&=\frac{\pal f}{\pal k}+\frac{\pal f}{\pal x}\,.\,\frac{\pal x}{\pal k}\\
	&=-\frac{1-nx}{n-k}+x\sqrt{\frac{1-kx}{(n-k)x}}\ln\left[\frac{1-kx}{(n-k)x}\right]\\
	&\leq -\frac{1-nx}{n-k}+x\left[\frac{1-kx}{(n-k)x}-1\right]\\
	&=0,
\end{align*}
where in the inequality we have used the fact that $\sqrt{y}\, \ln y\leq y-1$ for all $y\geq 1$. Thus, $f(x,k)$ is a decreasing function of $k$, and the minimum of $f$ is obtained for $k=n-1$. Finally, a global optimization over $n\leq d$ is required. Similar to the above method, we will show that for a fixed $g(n-1,x(n))$, $f(n-1,x(n))$ is a decreasing function in $n$. Solving $0=\frac{dg}{dn}=\frac{\pal g}{\pal n}+\frac{\pal g}{\pal x}.\frac{\pal x}{\pal n}$ ,
\[\frac{\pal x}{\pal n}=\frac{x\left(\sqrt{x}-2  \sqrt{1+x-n x}\right)}{(n-1) \left(\sqrt{1+x-n x}-\sqrt{x}\right)}.\] Substituting into $\frac{df}{dn}$, we get
\begin{align*}
	\frac{df}{dn}&=\frac{x \left[\sqrt{1+x-n x}-\sqrt{x}+\sqrt{1+x-n x} \ln \left(\frac{x}{1+x-n x}\right)\right]}{\sqrt{1+x-n x}-\sqrt{x}}\\
	&\leq\frac{x \left[\sqrt{1+x-n x}-\sqrt{x}+2\sqrt{1+x-n x} \left(\sqrt{\frac{x}{1+x-n x}}-1\right)\right]}{\sqrt{1+x-n x}-\sqrt{x}}\\
	&=-x,
\end{align*}
where in the inequality we have used the fact that $\ln y\leq y-1$ for all $y>0$. 
Thus, $f$ is a decreasing function of the dimension $d$, and hence the minimum of $f$ is attained at $n=d$. The minimum is obtained at $\lambda$ all of whose $(d-1)$ components are equal and the rest one is at least $1/d$. Assuming this larger component to be $\alpha$, the unique $\alpha\geq 1/d$ is obtained from the constraint $g(\alpha,d)=\sqrt{1+b}$ as
\[\alpha=\frac{2+(d-2)(d-b)+2 \sqrt{(b+1) (d-1) (d-1-b)}}{d^2}.\] For easy verification of the constraint, we note that 
\begin{align*}
	\sqrt{\alpha }&=\frac{\sqrt{(d-1) (d-1-b)}+\sqrt{b+1}}{d},\\
	\sqrt{1-\alpha }&=\frac{\sqrt{(b+1) (d-1)}-\sqrt{d-1-b}}{d}.
\end{align*}
This gives the lower bound of $C_r$ in Eq.~\eqref{Bound:tightest.possible.pure}. Since $f(x,k)$ is decreasing in $k$ and $n$, for a given $C_{\ell_1}=b$, if $n-2<b\leq n-1$ then the maximum of $f$ occurs inside $\Delta_n$ and the corresponding $\lambda$ will have one components $\beta\leq1/n$ and all the other $(n-1)$ components larger than $\beta$. Solving the constraint $g(x,n)=\sqrt{1+b}$ then gives the unique $\beta\leq 1/n$ as
\[\beta=\frac{2+(n-2)(n-b)-2 \sqrt{(b+1) (n-1) (n-1-b)}}{n^2}.\] 
We note that \begin{align*}
	\sqrt{\beta }&=\frac{\sqrt{b+1}-\sqrt{(n-1) (n-1-b)}}{n},\\
	\sqrt{1-\beta }&=\frac{\sqrt{(b+1) (n-1)}+\sqrt{n-1-b}}{n}.
\end{align*}\qed

\section{\label{App:Proof:Ge.upper.mixed} Proof of Theorem~\ref{Thm:Gen.Upper.mixed}}
Let $\tau\in\mathscr{I}$ be an optimal state for $C_R(\rho)$. Then from the definitions in Eq.~\eqref{Def:primary.quantities},
\begin{align*}
	C_r(\rho)&\leq S(\rho\|\tau)=\tr\left[\rho\left(\log_2\rho-\log_2\frac{(1+s)\tau}{(1+s)}\right)\right]\\
	&=\log_2(1+s)+\tr\left[\rho\left(\log_2\rho-\log_2(1+s)\tau\right)\right].
\end{align*}
Since $\rho\leq (1+s)\tau$ and $\log$ is operator monotone, the trace term is non-positive, and the first inequality follows. The last inequality follows from $C_R\leq C_{\ell_1}$ \cite{Napoli+5.PRL.2016,Piani+5.PRA.2016}, for which an independent proof is given below.

The dual of $C_R$ from Eq.~\eqref{Def:primary.quantities} gives \begin{align*}
	1+C_R(\rho)&=\max_{T\geq 0\,\&\,\diag(T)=\id}\,\tr[\rho T]\\&=\sum T_{ij}\rho_{ji}=1+2\sum_{i<j}\Re(T_{ij}\rho_{ji})\\
	&\leq 1+2\sum_{i<j}|T_{ij}\rho_{ji}|=1+C_{\ell_1},
\end{align*}
where we have used the inequality $|T_{ij}|\leq 1$ which follows from $\binom{1~~\bar{T}_{ij}}{T_{ij}~~1}\geq 0$.\qed

\section{\label{App:Support.4.Conj} Evidences for the conjecture $C_r(\rho)\leq C_{\ell_1}(\rho)$} 

Here we present two propositions to support our conjecture that $C_r(\rho)\le C_{\ell_1}(\rho)$ also holds for $C_{\ell_1}(\rho)\le 1$. Let us consider $ \rho= r+ \delta$, with diagonal $r$ and off-diagonal $\delta$, and the family of states $ \rho(\epsilon)=r+\epsilon\delta$ for $0\le \epsilon\le 1$. Then both $C_r[\rho(\epsilon)]$ and $C_{\ell_1}[\rho(\epsilon)]$ are analytic functions of $\epsilon$. 
\begin{proposition}
	\label{Prop:smallC_l_1} For a given $\rho$, consider the family of states $\rho(\epsilon)$. For $\epsilon\to 0$, \[C_r[\rho(\epsilon)]= O(\epsilon^2)\leq C_{\ell_1}[\rho(\epsilon)]=\epsilon C_{\ell_1}[\rho(1)]=O(\epsilon).\] 
	This shows that the conjecture 
	holds when the coherences are infinitesimally small.
\end{proposition}
\noindent Proof: We have $C_r[\rho(0)]=C_{\ell_1}[\rho(0)]=0$. Moreover, 
\[\frac{d}{d\epsilon} C_{\ell_1}[\rho(\epsilon)]\Bigr|_{\epsilon=0}=C_{\ell_1}[\rho(1)]>0,\] since we assume $\delta\neq 0$. Denoting the eigenvectors and eigenvalues 
of $\rho(\epsilon)$ by $|\lambda_i(\epsilon)\rangle$ and $\lambda_i(\epsilon)$  respectively, we have $C_r[\rho(\epsilon)]=H(r)-H[\lambda(\epsilon)]$. The Hellmann-Feynman theorem \cite{Hellmann.ZP.1933,Feynman.PR.1939} states
\begin{equation}
	\label{Eq:Hellmann-Feynman}
	\frac{d}{d\epsilon}\lambda_i(\epsilon)=\left\langle \lambda_i(\epsilon)\left|\frac{d}{d\epsilon}\rho(\epsilon)\right|\lambda_i(\epsilon)\right\rangle=\left\langle \lambda_i(\epsilon)\left|\delta\right|\lambda_i(\epsilon)\right\rangle,\;\forall i.
\end{equation}
Since $\tr[\rho(\epsilon)]=1$, we also have
\begin{equation}
	\label{Eq:Derivative.is.0}
	\sum_i \frac{d}{d\epsilon}\lambda_i(\epsilon)=0.
\end{equation}
Using Eqs.~(\ref{Eq:Hellmann-Feynman}) and \eqref{Eq:Derivative.is.0} we get
\begin{align*}
	\frac{d}{d\epsilon}C_r[\rho(\epsilon)]\Bigr|_{\epsilon=0}&=\sum_i \langle \lambda_i(0)|\delta|\lambda_i(0)\rangle \log_2[\lambda_i(0)]\\&=0.
\end{align*} \qed

We now give a rough bound on allowed $\epsilon$.
\begin{proposition}
	\label{Prop:moderatelC_l_1} If
	\[\int\limits_0^\epsilon  \log_2\left[\frac{\lambda_{\max}(\epsilon')}{\lambda_{\min}(\epsilon')}\right]\,d\epsilon'\leq  C_{\ell_1}[\rho(\epsilon)]=\epsilon C_{\ell_1}[\rho(1)],\]
	then $C_r[\rho(\epsilon')]\leq C_{\ell_1}[\rho(\epsilon')]$ for all $0\leq\epsilon'\leq \epsilon$.
	This shows that the conjecture 
	is correct for some states, even  if their $ C_{\ell_1}[\rho(\epsilon)]$ is smaller than one.
\end{proposition}
\noindent Proof: As in the previous proposition, 
\[\frac{d}{d\epsilon}C_r[\rho(\epsilon)]=\sum_i \langle \lambda_i(\epsilon)|\delta|\lambda_i(\epsilon)\rangle \log_2[\lambda_i(\epsilon)].\]
We observe that $\langle \lambda_i(\epsilon)|r+\delta|\lambda_i(\epsilon)\rangle\geq\lambda_{\min }(1)$ and $\log_2[\lambda_i(\epsilon)]\leq 0$, so that
\begin{align*}
	\frac{d}{d\epsilon}C_r[\rho(\epsilon)]&\leq\sum_i \Big[\lambda_{\min}(1)-\langle \lambda_i(\epsilon)|r|\lambda_i(\epsilon)\rangle\Big] \log_2[\lambda_i(\epsilon)]\\
	&\leq d\lambda_{\min}(1)\,\log_2[\lambda_{\max}(\epsilon)]\\
	&\qquad\qquad-\sum_i\langle\lambda_i(\epsilon)|r|\lambda_i(\epsilon)\rangle\,\log_2[\lambda_{\min}(\epsilon)]\\
	&\leq \log_2\left[\frac{\lambda_{\max}(\epsilon)}{\lambda_{\min}(\epsilon)}\right]\label{Eq:M.Bound}\numberthis.
\end{align*}
The proposition follows by integrating over $\epsilon'$ from zero to $\epsilon$. \qed 

\noindent\textbf{Examples}: To illustrate usefulness of the above propositions, let us consider families of states with $r=\id/d$. Then 
\begin{align*}
	\lambda_{\max}(\epsilon)&=\max_{|\psi\rangle}\:\langle \psi| (1-\epsilon)r + \epsilon (r+\delta)|\psi\rangle \\ &=
	(1-\epsilon)/d + \epsilon \lambda_{\max}(1)\\ &\le \lambda_{\max}(1),
\end{align*}as $1/d\leq\lambda_{\max}(1)$. Similarly, $\lambda_{\min}(\epsilon)
\geq \lambda_{\min}(1)$. Substituting the bounds in Eq.~\eqref{Eq:M.Bound} and integrating over $\epsilon'$ from zero to $\epsilon$, we get $C_r[\rho(\epsilon)]\le \epsilon \log_2[\lambda_{\max}(1)/\lambda_{\min}(1)]$. Thus,
\begin{align*}
	&C_{\ell_1}[\rho(1)]\ge \log_2\left[\frac{\lambda_{\max}(1)}{\lambda_{\min}(1)}\right]\numberthis\label{Eq:M.cond.log.M/m}\\
	\Rightarrow\: & C_r[\rho(\epsilon)]\le C_{\ell_1}[\rho(\epsilon)],\: \forall 0\leq\epsilon\leq 1.
\end{align*}
The set of states fulfilling condition~\eqref{Eq:M.cond.log.M/m} is not empty -- in particular all matrices for which
$ \log_2[\lambda_{\max}(1)/\lambda_{\min}(1)]\le 1$, i.e.  $\lambda_{\rm max}(1)/\lambda_{\rm min}(1)\le 2$ typically fulfill
this condition if $C_{\ell_1}(\rho(1))$ is not too large. Examples of such matrices are easily constructed with all off-diagonal elements equal.

\section{\label{App:Prop:Ultimate.bounds}Proof of Theorem~\ref{Prop:Ultimate.bounds}} 
First note that the following two optimization problems are equivalent (solving one equally solves the other):
\begin{align}
	t(b)&=\min C_r(\rho)\quad\text{s.t. } C_R(\rho)\geq b\label{Eq:Opt:min.cr.cl1.geq.b}\\
	b(t)&=\max C_R(\rho)\quad\text{s.t. } C_r(\rho)\leq t\label{Eq:Opt:max.cl1.cr.leq.t}
\end{align}
Due to convexity of the functions involved, in each case the optimum will occur on the equality condition.

Now, using dual form of $C_R$ \cite{Napoli+5.PRL.2016,Piani+5.PRA.2016},
\begin{align*}
	1+C_R(\rho)&=\max \tr[D] \quad\text{ s.t. } \rho\leq D\, \&\,D=\diag(D)\\
	&=\max \tr[\rho B] \quad\text{ s.t. } B\geq 0\, \&\, \diag(B)=\id\\
	&\geq \tr[\rho J],\label{Eq:cl1.geq.tr.rhoJ}\numberthis
\end{align*} where $J=d|\Psi\ran\lan\Psi|$ is the matrix having all entries $1$. 
Invoking Eqs.~\eqref{Eq:Opt:min.cr.cl1.geq.b} and \eqref{Eq:Opt:max.cl1.cr.leq.t}, our problem reduces to finding
\begin{equation}
	\label{Eq:max.tr.rhoJ.s.t.Cr.leq.t} 1+ b(t)\ge \max\,\tr[\rho J]\quad\text{s.t. }C_r(\rho)\leq t.
\end{equation}
Note, that both quantities  $\tr[\rho J]$ and $C_r(\rho)$ in Eq.~\eqref{Eq:max.tr.rhoJ.s.t.Cr.leq.t} remain invariant under permutations of rows and columns of $\rho$. Replacing \[\rho \longmapsto\frac{1}{d!}\sum_{\pi}U_{\pi}\rho U_{\pi}^\dagger,\]
where the sum runs over the permutations, shows that the maximum is achieved at $\rho=p|\Psi\ran\lan\Psi|+(1-p)\id/d$. The value of $p$ is determined by the condition $C_R=C_{\ell_1}(\rho)=b$.\qed

Our numerical study indicates that the same state may have minimum $C_r$ for a given $C_{\ell_1}$ as well. Unfortunately, the above method is not applicable  in this case.

\section{$C_{\ell_1}$ could be smaller than its convex roof?} 
\label{App:Failed.Attempt} 

As usual, let us define the convex roof extension of $C_{\ell_1}$ as (see also \cite{Qi+2.JPA.2017,Chin.PRA.2017})
\begin{equation}\label{Eq:convex.roof.Cl1}
	C_1(\rho):=\min_{\{p_i,\,|\psi_i\ran\}}\left\{\sum_ip_iC_{\ell_1}(|\psi_i\ran)\, \big\rvert\,\rho=\sum_ip_i|\psi_i\ran\lan\psi_i|\right\}.
\end{equation}
Then $C_{\ell_1}(\rho)\leq C_1(\rho)$. Note that if equality occurs for a state $\rho$, then by convexity of $C_r$ and Proposition~\ref{Prop:Cr=Cl1.iff}, it must satisfy  $C_r(\rho)\leq C_{\ell_1}(\rho)$. The result below gives another proof that all qubits fulfill our conjecture.

\begin{proposition}
	All qubit states $\rho$ satisfy $C_{\ell_1}(\rho)=C_1(\rho)$.
\end{proposition}

\noindent Proof: It suffices to  show that for given $a,b$ there are $p,\lambda,\mu,$ such that
\begin{multline*}
	\begin{pmatrix}
		a & b\\b & 1-a
	\end{pmatrix}=p\begin{pmatrix}
		\lambda&\sqrt{\lambda(1-\lambda)}\\
		\sqrt{\lambda(1-\lambda)} & 1-\lambda
	\end{pmatrix}\\+(1-p)\begin{pmatrix}
		\mu & \sqrt{\mu(1-\mu)}\\\sqrt{\mu(1-\mu)} & 1-\mu
	\end{pmatrix},\end{multline*}
where we may assume without loss of generality $a\in(0,1/2]$, $b\in(0,1/2)$, with $a(1-a)>b^2$. This (positivity) demands that $(1-\sqrt{1-4b^2})/2<a\leq1/2$. Since we require each of the pure states to have $C_{\ell_1}=2b$, $\lambda,\mu$ are necessarily the two roots of $x(1-x)=b^2$. Setting $\lambda=(1+\sqrt{1-4b^2})/2$, $\mu=1-\lambda$, and comparing the first diagonal element we get a unique solution
\[p=\frac{1}{2}-\frac{1}{2}\frac{1-2a}{\sqrt{1-4b^2}}.\]
Since $(1-2a)^2=1-4a(1-a)\leq 1-4b^2$, hence $p\in(0,1/2]$. Note that the standard spectral decomposition does not help, as each of the eigen-projectors have 
$C_{\ell_1}=2b/\sqrt{1-4[a(1-a)-b^2]}>2b$.\qed

As mentioned in the proof of Theorem~\ref{Thm:Gen.Upper.mixed}, $C_{\ell_1}(\rho)$ is exactly double of the negativity of the corresponding maximally correlated state $\sigma$:  
\begin{align*}C_{\ell_1}\left(\rho\right)&=2\N\left(\sigma\right),\\
	\text{where }\:\rho&=\sum a_{ij}|i\ran\lan j|,\\
	\text{and }\:\sigma& =\sum a_{ij}|ii\ran\lan jj|.
\end{align*}
Therefore, denoting the convex roof of negativity by $\N_c$, 
\begin{align}
	&C_{\ell_1}(\rho)=C_1(\rho)\label{Eq:N=Nc?1}\\ \Leftrightarrow\: &\N(\sigma)=\N_c(\sigma)\label{Eq:N=Nc?2}
\end{align}
It is known \cite{Lee+3.PRA.2003} that equality occurs in Eq.~\eqref{Eq:N=Nc?2} for isotropic states while strict inequality for Werner states (in $d>2$). Unfortunately, none of those states is maximally correlated for $d>2$; hence we cannot browse the results directly into the coherence scenario. It was observed in Ref.~\cite{Qi+2.JPA.2017} that equality holds in Eq.~(\ref{Eq:N=Nc?1}) for all pseudopure states defined in Proposition~\ref{Prop:Case.of.pseudo.mixture}. Also, it was shown in \cite[Theorem~3]{Chin.PRA.2017} that strict inequality occurs for a similar quantity.
\end{document}